\documentstyle[sprocl,epsfig]{article}
 
\def\st{\mbox{$\rm \tilde{t}_1$}}
\def\s2{\mbox{$\rm \tilde{t}_2$}}
\def\stl{\mbox{$\rm \tilde{t}_L$}}
\def\sr{\mbox{$\rm \tilde{t}_R$}}
\def\mt{\mbox{$m_{\rm \tilde{t}_1}$}}
\def\m2{\mbox{$m_{\rm \tilde{t}_2}$}}
\def\cost{\mbox{$\cos\theta_{\rm \tilde t}$}}
\def\sint{\mbox{$\sin\theta_{\rm \tilde t}$}}
\def\ee{\mbox{$\rm e^+e^-$}}

\begin{document}
\begin{titlepage}
\def\thefootnote{\fnsymbol{footnote}} 
 
\begin{center}
\mbox{ }
 
\vspace*{-3cm}
 
 
\end{center}
\begin{flushright}
\Large
\mbox{\hspace{10.2cm} hep-ph/0102341} \\
\mbox{\hspace{10.2cm} IEKP-KA/2001-6} \\
\mbox{\hspace{10.2cm} February 2001}
\end{flushright}
\begin{center}
\vskip 2.0cm
{\Huge\bf
Study of Scalar Top Quarks in the
\vskip 0.2cm
Neutralino and Chargino Decay
\vskip 0.4cm
Channel}
\vskip 2.5cm
{\LARGE\bf H.~Nowak$^1$ and A.~Sopczak$^2$}\\
\smallskip
\vskip 1cm
\Large $^1$ DESY Zeuthen, $^2$ University of Karlsruhe
 
\vskip 2.5cm
\centerline{\Large \bf Abstract}
\end{center}

\vskip 1.5cm
\hspace*{-1cm}
\begin{picture}(0.001,0.001)(0,0)
\put(,0){
\begin{minipage}{16cm}
\Large
\renewcommand{\baselinestretch} {1.2}
The scalar top discovery potential has been studied
with a full-statistics background simulation at $\sqrt{s}=500$~GeV
and ${\cal L}=500$~fb$^{-1}$ for the TESLA project.
The beam polarization is very important to measure the scalar top 
mixing angle and to determine its mass.
The latest estimation of the beam polarization parameters is applied.
This study includes $\rm e^+$ polarization, which improves the
sensitivity.
For a 180 GeV scalar top at minimum production cross section, 
we obtain 
$\Delta m=0.8$~GeV and $\Delta\cost=0.008$ in the
neutralino decay channel,
and $\Delta m=0.5$~GeV and $\Delta\cost=0.004$
in the chargino decay channel.

\renewcommand{\baselinestretch} {1.}

\normalsize
\vspace{2.cm}
\begin{center}
{\sl \large
Talk at the Worldwide Workshop on Future $\rm e^+e^-$ Collider,
Chicago, November 2000, \\ to be published in the proceedings.
\vspace{-6cm}
}
\end{center}
\end{minipage}
}
\end{picture}
\vfill
 
\end{titlepage}
 
 
\newpage
\thispagestyle{empty}
\mbox{ }
\newpage
\setcounter{page}{1}

\title{Study of Scalar Top Quarks in the
Neutralino and Chargino Decay Channel}

\author{H.~Nowak$^1$ and A.~Sopczak$^2$\footnote{speaker}}
\address{\it $^1$ DESY Zeuthen, $^2$ University of Karlsruhe}

\maketitle

\abstracts{
The scalar top discovery potential has been studied
with a full-statistics background simulation at $\sqrt{s}=500$~GeV
and ${\cal L}=500$~fb$^{-1}$ for the TESLA project.
The beam polarization is very important to measure the scalar top 
mixing angle and to determine its mass.
The latest estimation of the beam polarization parameters is applied.
This study includes $\rm e^+$ polarization, which improves the
sensitivity.
For a 180 GeV scalar top at minimum production cross section, 
we obtain 
$\Delta m=0.8$~GeV and $\Delta\cost=0.008$ in the
neutralino decay channel,
and $\Delta m=0.5$~GeV and $\Delta\cost=0.004$
in the chargino decay channel.}

\vspace*{-0.8cm}
\section*{Introduction}
The study of the scalar top quarks is of particular interest, since the
lighter stop mass eigenstate is likely to be the lightest scalar quark in a 
supersymmetric theory. The mass eigenstates are \mt\ and \m2\ with 
$\mt < \m2$, where $\st=\cost \stl + \sint \sr$ and 
$\s2=-\sint \stl + \cost \sr$ with the mixing angle \cost.
We study the experimental possibilities to
determine \mt\ and \cost\ at a high-luminosity \ee\ linear collider 
like the TESLA project~\cite{tesla} with polarized
$\rm e^+$ and $\rm e^-$ beams.

The simulated production process is 
$\ee\rightarrow\st\bar{\st}$ with two decay modes
$\st\rightarrow\tilde{\chi}^0$c and 
$\st\rightarrow\tilde{\chi}^+$b.
A 100\% branching fraction  in each decay mode is simulated.
The first scalar top decay into a c-quark and the lightest neutralino 
results in a signature of two jets and large missing energy.
The second investigated stop decay mode leads also to large missing energy
and further jets from the chargino decay.
The neutralino channel is dominant unless the decay into a chargino is
kinematically allowed.
Details of the event simulation with SGV~\cite{sgv} tuned for a 
TESLA detector~\cite{tesla} are given in Ref.~\cite{epj}.
The signals and a total of 16 million Standard Model 
background events are simulated (Table~\ref{tab:pre})
for ${\cal L}=500$~fb$^{-1}$.

\begin{table}[hp]
\begin{center}
\begin{tabular}{|c|c|c|c|c|c|c|c|c|}\hline
Channel      & $\tilde{\chi}^0$c$\tilde{\chi}^0\mathrm{\bar{c}}$
             & $\tilde{\chi}^+$b$\tilde{\chi}^-\mathrm{\bar{b}}$
             &eW$\nu$  & WW & $\rm q\bar{q}$   
             & t$\rm \bar{t}$&ZZ   & eeZ   \\ \hline
(in 1000)    & 50   & 50 & 2500 & 3500  &6250  & 350 & 300 & 3000 \\ \hline
\end{tabular}
\end{center}
\caption{\label{tab:pre} 
Number of simulated signal and background events.}
\end{table}

\vspace*{-0.5cm}
\section*{Neutralino Channel}
The reaction $\rm \ee\rightarrow\st\bar{\st}\rightarrow 
\tilde{\chi}^0c\tilde{\chi}^0\bar{c}
$ has been studied for a 180 GeV scalar top and a 100 GeV neutralino.
After a preselection,
278377 background events remain~\cite{sitgesproc,epj}.
In order to separate the signal from the background, 
the following selection variables are defined: 
visible energy,
number of jets, 
thrust value and direction, 
number of clusters, 
transverse and parallel imbalance, 
acoplanarity and invariant mass of two jets~\cite{sitgesproc,epj}.
An Iterative Discriminant Analysis (IDA)~\cite{ida} optimized 
the selection. For unpolarized beams and 
12\% efficiency, 400 background events are expected.

\begin{figure}[hb]
\begin{center}
\mbox{\epsfig{file=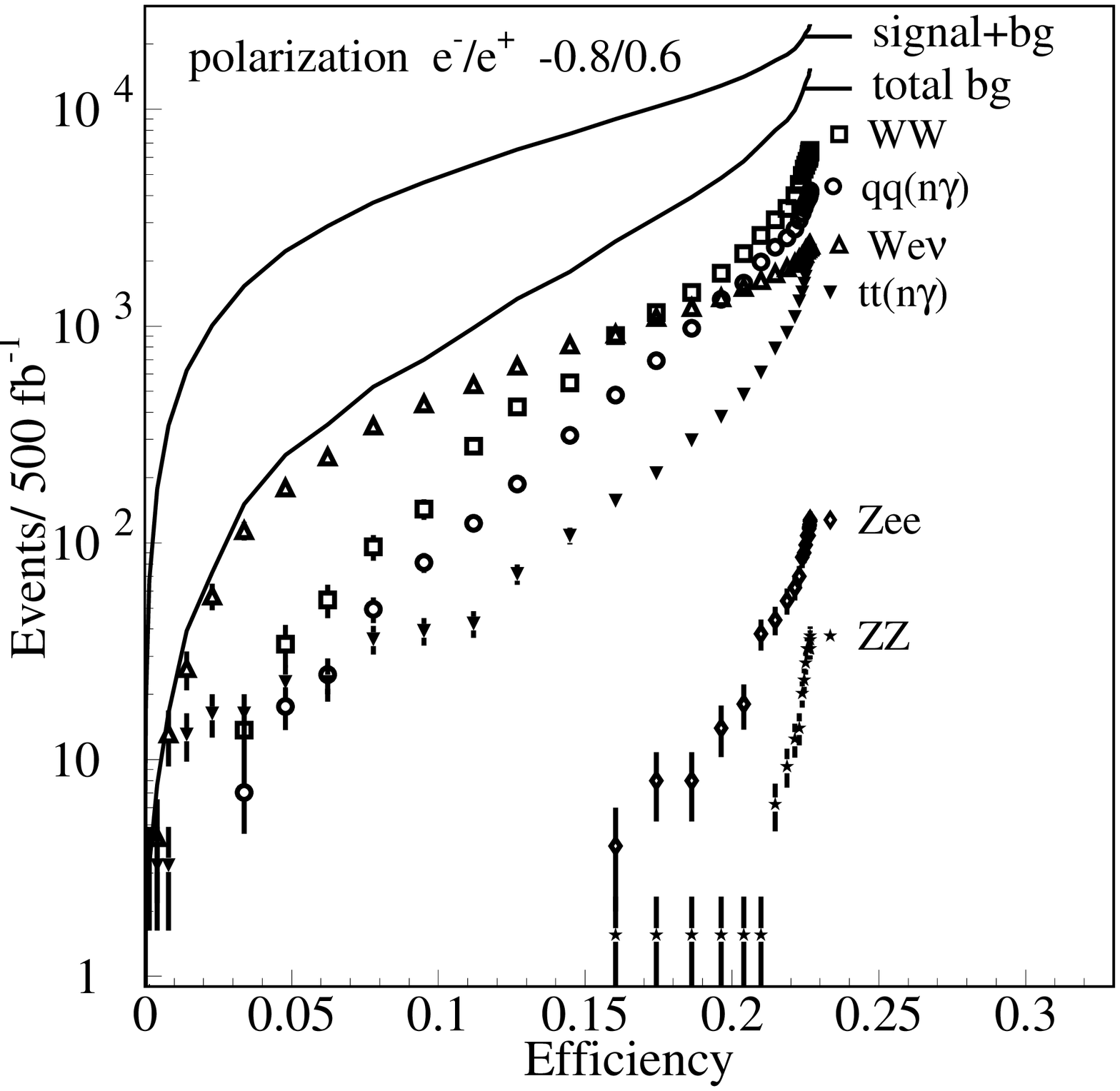,width=0.49\textwidth}}
\mbox{\epsfig{file=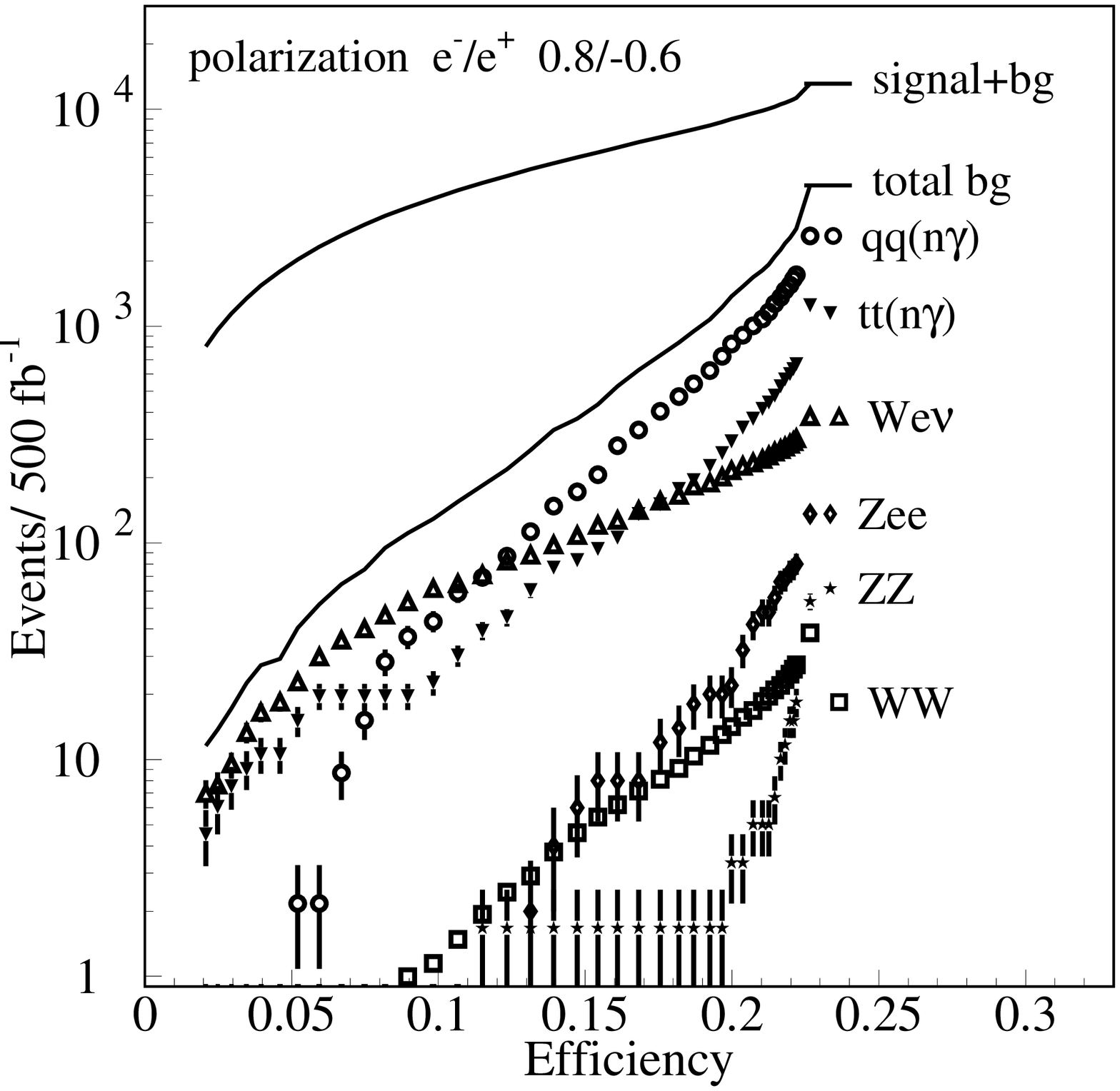,width=0.49\textwidth}}
\end{center}
\caption{\label{fig:pol} 
Background vs. signal efficiency for left- and right-polarization
in the neutralino decay channel
for a 180 GeV scalar top and a 100 GeV neutralino.}
\end{figure}

The polarization of the $\rm e^+$ and $\rm e^-$ beams at a future 
linear collider offers the opportunity to enhance or suppress 
the left- or right-handed couplings 
of the scalar top signal and to determine mass and 
mixing angle independently. The production cross section of each background 
process depends differently on the polarization. It is therefore important 
for a high-statistics analysis to study the expected background channels 
individually.
The expected cross sections~\cite{xsec,generator} are given in 
Table~\ref{tab:bgxsec} for different beam polarization states.
The IDA analysis was repeated for $-0.9$ and $0.9$ 
polarization~\cite{epj}\footnote{For a polarization of 
$-0.9$, 95\% of the $\rm e^-$ are left-polarized.
In the previous analyses~\cite{moriokaproc,desy123d,desy123e,zphys}
it was assumed that only 90\% of the $\rm e^-$ were polarized.}.
Here, we recalculate all rates for the new machine polarization.
Figure~\ref{fig:pol} shows the number of background events
as a function of the signal efficiency for  
$-0.8 / 0.6$ (left) and $0.8 / -0.6$ (right) polarization.
For 12\% detection efficiency, 1194 background events are expected
leading to $\sigma_{\rm left} = 81.8 \pm 1.3 $ fb, 
and 208 background events giving $\sigma_{\rm right} = 76.4 \pm 1.2 $ fb, 
where $\Delta\sigma/ \sigma = 
\sqrt{N_{\rm signal} + N_{\rm background}} / N_{\rm signal}$.

\clearpage
\begin{table}[tp]
\vspace*{-0.5cm}
\begin{center}
\begin{tabular}{|c||c|c|c|c|c|c|c|} \hline
 Pol.  &Pol. & $\st\bar{\st}$ &$\rm W e \nu$ &  WW & $\rm q\bar q$ &$\rm t\bar t$ &  ZZ  \\
 of $\rm e^-$ & of $\rm e^+$ & CALVIN
&{\small \hspace*{-0.5mm}GRACE}\hspace*{-0.5mm}     
&{\small \hspace*{-0.5mm}WOPPER}\hspace*{-0.5mm}   
&{\small \hspace*{-0.5mm}HERWIG}\hspace*{-0.5mm} 
&{\small \hspace*{-0.5mm}HERWIG}\hspace*{-0.5mm}
&{\small \hspace*{-0.5mm}COMPHEP}\hspace*{-0.5mm} \\ \hline
  $-0.8$& 0.6    & 0.0818 & 10.7  & 22.6 & 21.5  & 1.11 & 0.909 \\
  $-0.9$& 0.0    & 0.0552 & 6.86  & 14.9   & 14.4  & 0.771  & 1.17  \\
  0     & 0.0    & 0.0535 & 5.59  &  7.86  & 12.1  & 0.574  & 0.864 \\
  0.9   & 0.0    & 0.0517 & 4.61  &  0.906 &  9.66 & 0.376  & 0.554 \\ 
  0.8   & $-0.6$ & 0.0764 & 1.78 & 0.786 & 13.0 & 0.542 & 0.464  \\ \hline
\end{tabular}
\end{center}
\caption{\label{tab:bgxsec} 
Signal and background cross sections (pb) from different
event generators for $\rm e^-$ and $\rm e^+$ polarization states
for $\mt=180$~GeV and $\cost=0.57$.
The Zee cross section is 0.6~pb.}
\end{table}

\section*{Chargino Channel}
\begin{figure}[ht]
\begin{center}
\vspace*{-1.5cm}
\mbox{\epsfig{file=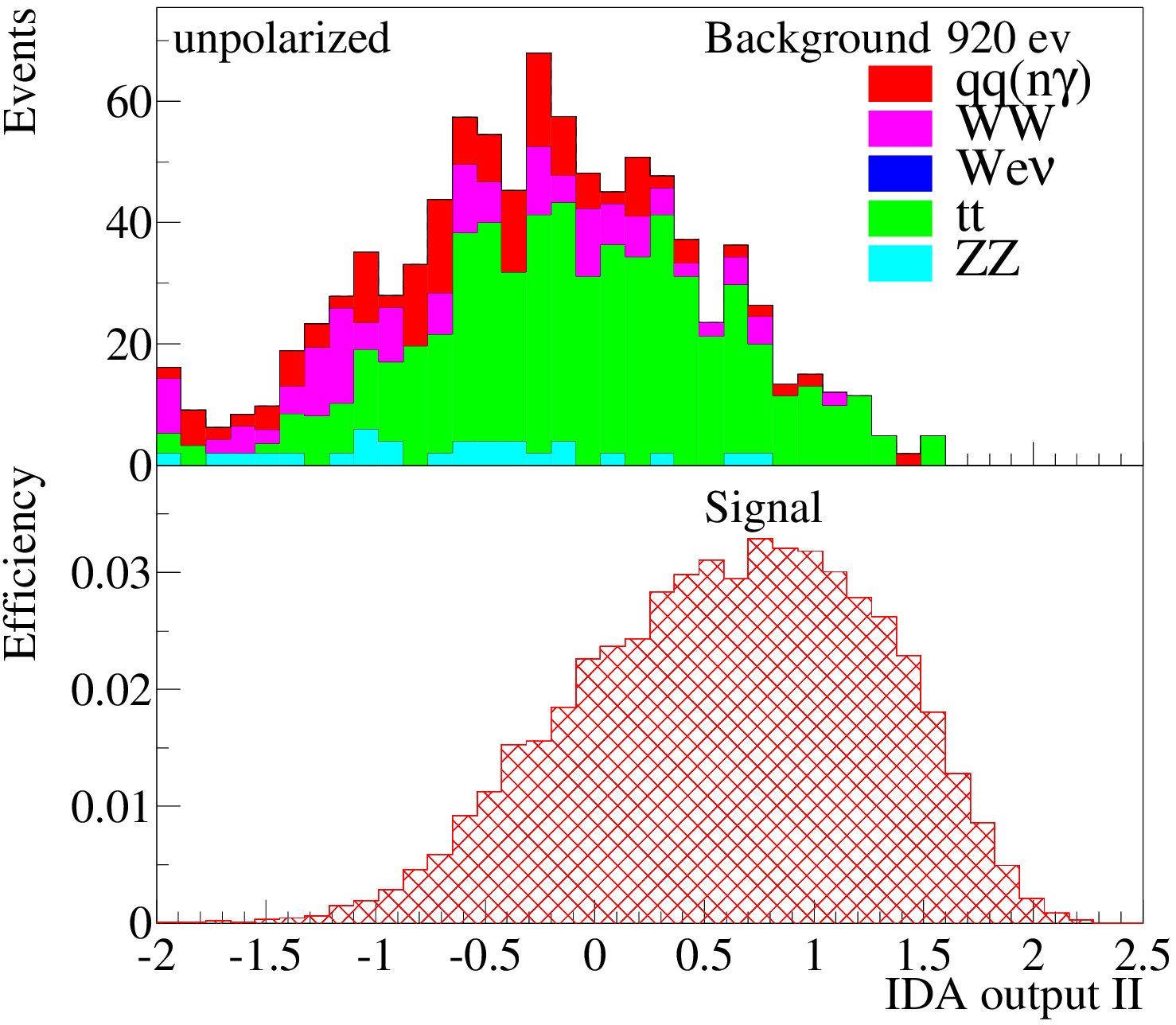,width=0.49\textwidth}}
\mbox{\epsfig{file=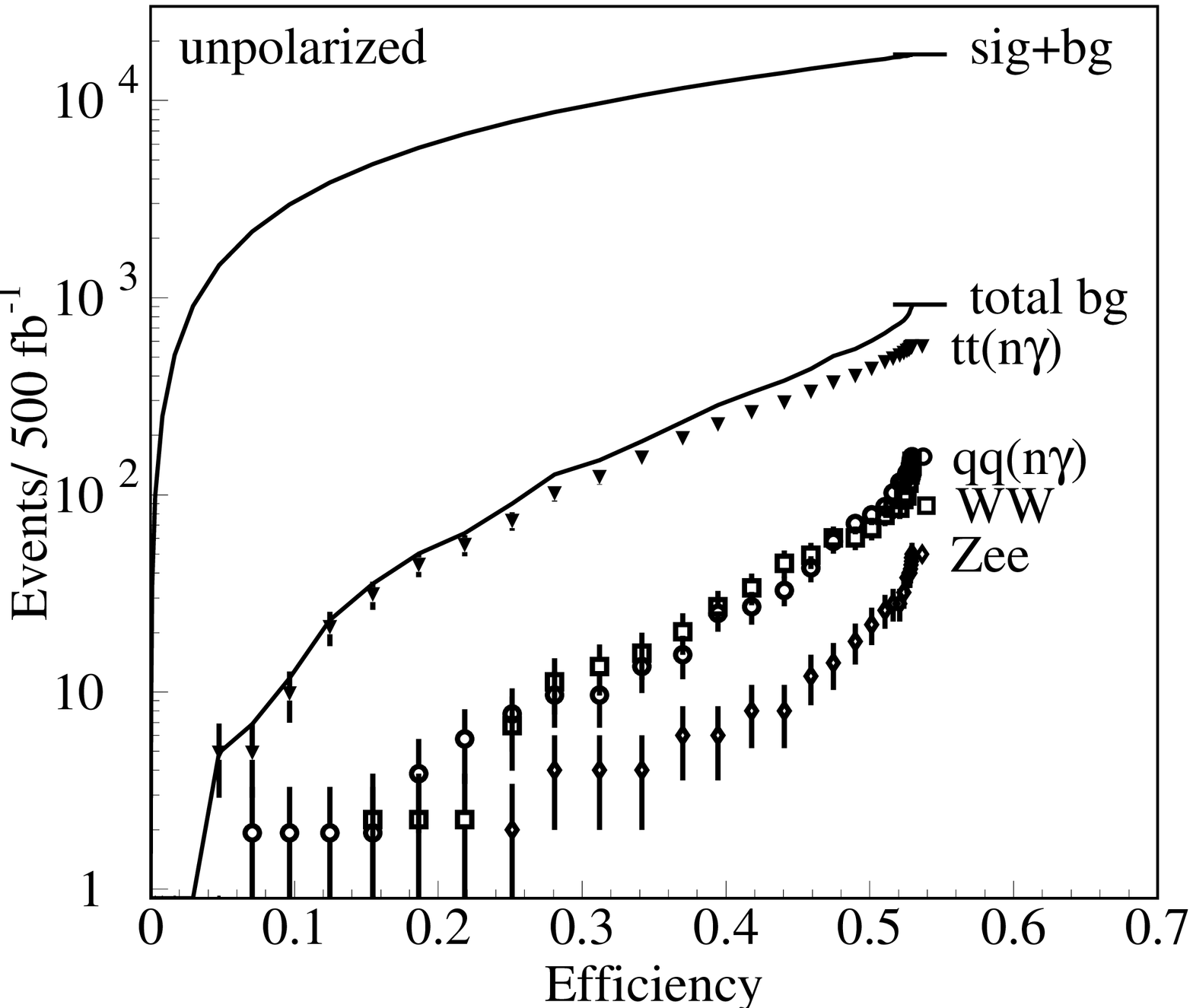,width=0.49\textwidth}}
\end{center}
\caption{\label{fig:ida2x} 
Final IDA output and background vs. signal efficiency 
in the chargino decay channel for unpolarized beams
for a 180 GeV scalar top, a 150 GeV chargino and a 60 GeV neutralino.}
\end{figure}

The reaction $\rm \ee\rightarrow\st\bar{\st}\rightarrow 
\tilde{\chi}^+b\tilde{\chi}^-\bar{b} \rightarrow
\tilde{\chi}^0W^+b\tilde{\chi}^0W^-\bar{b}$ 
has been studied with focus on the hadronic W decays
for a 180 GeV scalar top, a 150 GeV chargino and a 60 GeV neutralino, 
where the chargino decays 100\% into a W boson and a neutralino.
A preselection  similar to that for the study of the neutralino channel
is applied and 209051 background events remain~\cite{epj}.
In order to separate the signal from the background, 
the following selection variables are defined: 
visible energy,
number of jets,
thrust value,
number of clusters,
transverse and parallel imbalance,
and the isolation angle of identified leptons.
For an unpolarized beam, Fig.~\ref{fig:ida2x} shows the final IDA 
output variable and the resulting number of background events as a 
function of the signal efficiency.
For 12\% efficiency, only 20 background events are expected.
Allowing a background rate of 400 events, as in the neutralino channel,
the efficiency is 44\%, from which we derive the relative error
on the cross section to be 0.75\%.
In this case the number of expected signal events is much larger
than the expected background, thus no separate tuning of the IDA for
left- and right-polarization is required.
The background is neglected for the determination of mass and mixing 
angle. Note that the dependence of the background rates on the polarization
has to be taken into account for stop masses closer to the kinematic 
threshold.

\clearpage
\section*{Results}
\begin{figure}[hb]
\begin{center}
\mbox{\epsfig{file=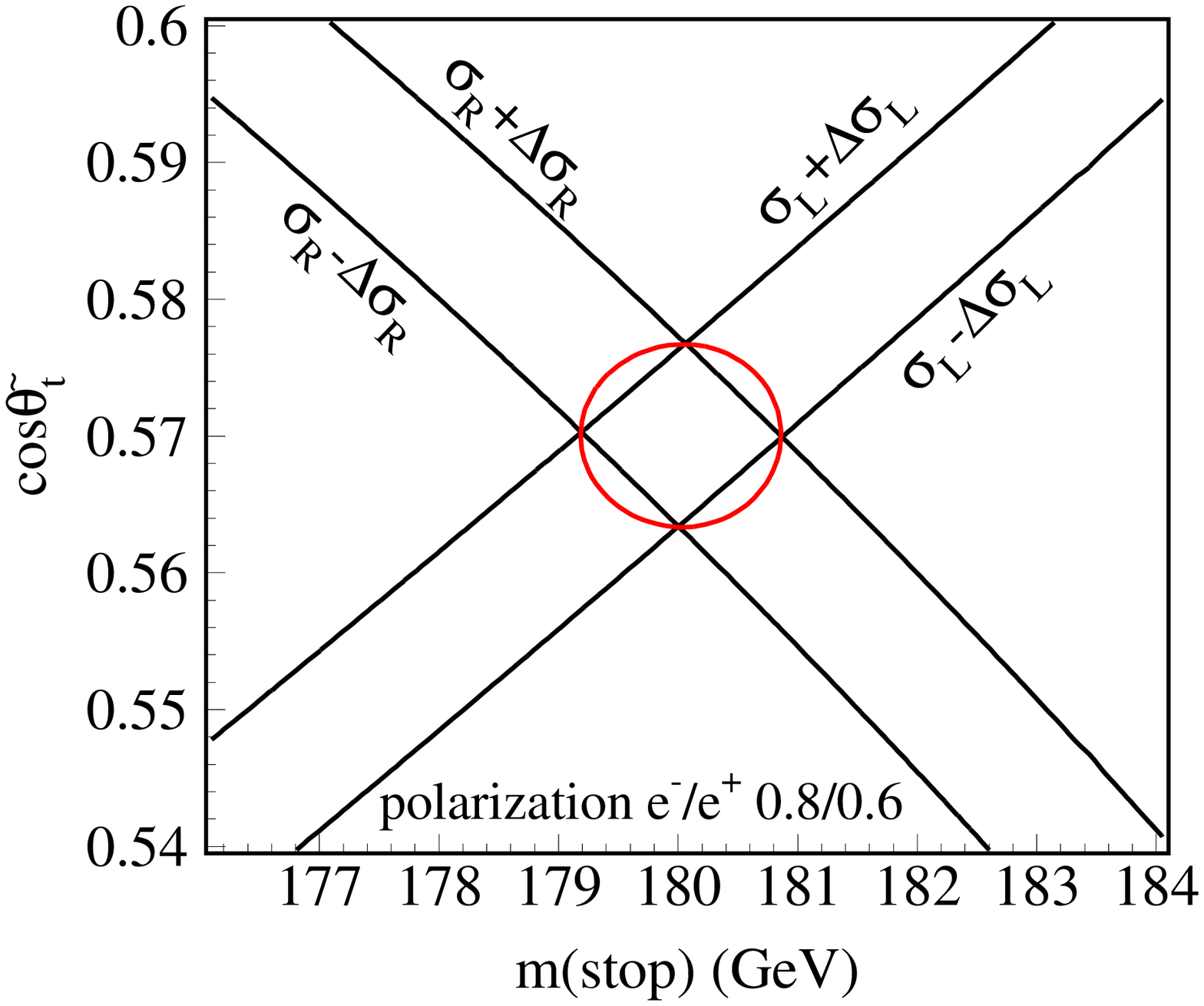,width=0.49\textwidth}}
\mbox{\epsfig{file=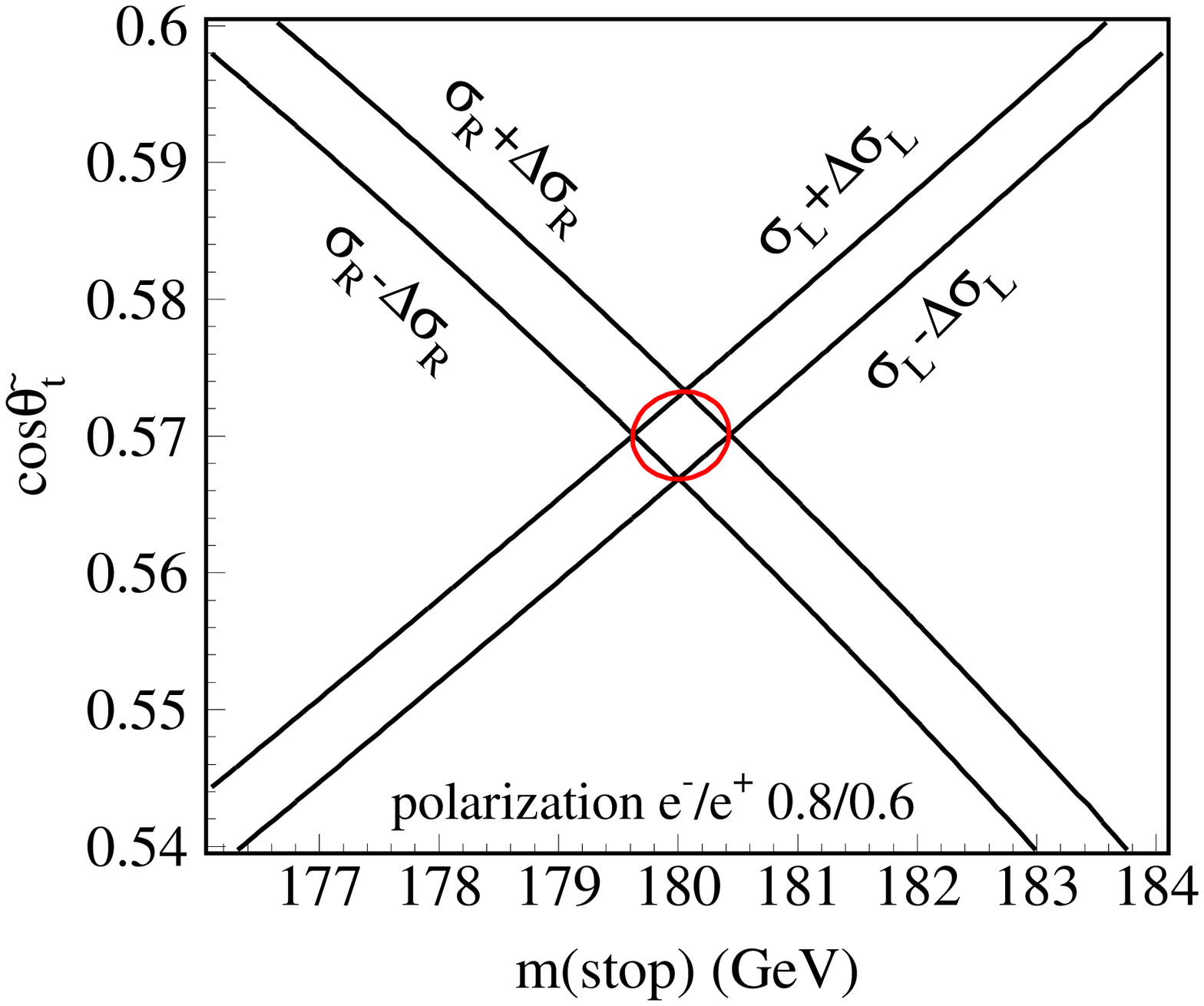,width=0.49\textwidth}}
\end{center}
\vspace*{-0.5cm}
\caption{\label{fig:ellipse} 
Error bands and the corresponding error ellipse 
as a function of \mt\ and \cost\ for $\sqrt s =500$~GeV
and ${\cal L}=500$~fb$^{-1}$. 
The dot corresponds to $\mt=180$~GeV and $\cost=0.57$.}
\vspace*{-3.5cm}
\end{figure}

For the neutralino and chargino decay channel of scalar top quarks,
we have determined the expected Standard Model background rate 
as a function of the signal efficiency. 
The total simulated background of about 16 million events is largely reduced,
which allows a precision measurement of the scalar 
top production cross section with a relative error of better than 2\%
in the neutralino channel and about 1\% in the chargino channel.
Based on experiences gained at LEP, we expect that detection efficiencies 
for other mass combinations are similar as long as the mass difference 
between the scalar top and the neutralino is larger than about 20 GeV.
Figure~\ref{fig:ellipse} shows the corresponding error bands and the error
ellipse in the \mt\ -- \cost\ plane for both decay channels
for $0.8/0.6$ left- and right-polarization of the $\rm e^-/e^+$ beams
and a luminosity of 500 fb$^{-1}$ each.
The statistical errors are a factor 7 better in the neutralino channel and
about a factor 14 better in the chargino channel than reported
previously~\cite{moriokaproc,desy123d,desy123e,zphys}, and improve further 
when in addition e$^+$ polarization is included. 
Detailed results are given in Table~\ref{tab:sum}.
The highest statistical precision is obtained in the chargino channel 
with an error $\Delta\mt=0.4$~GeV for $\mt=180$~GeV and
$\Delta\cost=0.003$ $\cost=0.570$.
Based on the experience from direct searches at LEP, the systematic errors
on the event selection are less than 1\%; precise investigations require
the detailed detector layout and a full simulation.
The stop generator~\cite{asgenerator} has been interfaced 
with the SIMDET~\cite{simdet} 
simulation to allow an independent test of the detector simulation and related
systematic errors.
Another uncertainty could arise from the luminosity measurement, 
the measurement of the polarization, 
and the theoretical uncertainty of the production cross section, 
which will be determined in the future.
A high-luminosity linear collider with the capability of beam 
polarization has a great potential for precision measurements 
in the scalar quark sector predicted by Supersymmetric theories.

\clearpage

\begin{table}[hp]
\vspace*{0.5cm}
\begin{center}
\begin{tabular}{|c|c|c||cc|cc|} \hline
Luminosity (fb$^{-1}$) & e$^-$ Pol. & e$^+$ Pol. 
& (a)  $\Delta\mt$ & $\Delta\cost$ & (b) $\Delta\mt$ & $\Delta\cost$ \\\hline
10  & 0.8  &  0.0  &  7.0 & 0.06  & 7.0  & 0.06  \\
500 & 0.9  &  0.0  &  1.0 & 0.009 & 0.5  & 0.004 \\
500 & 0.8  &  0.6  &  0.8 & 0.008 & 0.4  & 0.003 \\\hline
\end{tabular}
\end{center}
\caption{\label{tab:sum}
Expected errors on the scalar top mass and mixing angle from simulations
with different luminosity and beam polarization
in the neutralino (a) and chargino (b) channels.
The 10 fb$^{-1}$ 
analysis~\protect\cite{moriokaproc,desy123d,desy123e,zphys}
used a sequential event selection; while the 500~fb$^{-1}$ results in the
neutralino~\protect\cite{sitgesproc} and 
chargino~\protect\cite{epj} channels
are based on an IDA. The new result includes e$^+$ polarization.}
\vspace*{-1.0cm}
\end{table}

\end{document}